\documentclass[conference,10pt]{IEEEtran}
\usepackage{cite}
\usepackage{comment}
\usepackage{amsmath,amssymb,amsfonts}
\usepackage{algorithmic}
\usepackage{algorithm}
\usepackage{graphicx}
\usepackage{textcomp}
\usepackage{xcolor}
\usepackage{balance}
\usepackage{url}
\usepackage{subfigure}
\begin{document}

% Example definitions.
% --------------------
\def\b{{\mathbf b}}
\def\c{{\mathbf c}}
\def\n{{\mathbf n}}
\def\v{{\mathbf v}}
\def\x{{\mathbf x}}
\def\y{{\mathbf y}}
\def\X{{\mathbf X}}
\def\Y{{\mathbf Y}}
\def\F{{\mathbf F}}
\def\G{{\mathbf G}}
\def\H{{\mathbf H}}
\def\W{{\mathbf W}}
\def\A{{\mathbf A}}
\def\B{{\mathbf B}}
\def\C{{\mathbf C}}
\def\M{{\mathbf M}}
\def\Z{{\mathbf Z}}
\def\U{{\mathbf U}}
\def\V{{\mathbf V}}
\def\I{{\mathbf I}}
\def\w{{\mathbf w}}
\def\S{{\mathbf \Sigma}}
\def\SigmaW{{\mathbf \sigma_\v}}
\def\L{{\cal L}}

\newcommand{\temp}[1]{\textcolor{red}{#1}}
\newcommand{\mario}[1]{\textcolor{green}{#1}}

\IEEEoverridecommandlockouts

\title{Semantic Waveforms for AI-Native 6G Networks}
% 1\textsuperscript{st} 
\author{Nour Hello$^1$, Mohamed Amine Hamoura$^1$, Francois Rivet$^{2}$, and Emilio Calvanese Strinati$^1$ \medskip \\
$^1$ CEA Leti, University Grenoble Alpes, 38000, Grenoble, France.\\
$^2$ University of Bordeaux, CNRS, Bordeaux INP, IMS, UMR 5218, F-33400 Talence, France. 

\thanks{This work has been supported by the SNS JU project 6G-DISAC under the EU’s Horizon Europe research and innovation program under Grant Agreement No 101139130.} \vspace{-.2cm}}
%\vspace{-2cm}
%\author{\IEEEauthorblockN{Anonymous Authors}} % TODO
\maketitle

\begin{abstract} In this paper, we propose a semantic-aware waveform design framework for AI-native 6G networks that jointly optimizes physical layer resource usage and semantic communication efficiency and robustness, while explicitly accounting for the hardware constraints of RF chains.
% In this paper, we propose a semantic-aware waveform design framework for AI-native 6G networks that accounts for specific hardware limitations of RF chains, jointly optimizing physical layer resource usage with semantic communication efficiency and robustness.
Our approach, called Orthogonal Semantic Sequency Division Multiplexing (OSSDM), introduces a parametrizable, %Walsh-based orthogonal 
orthogonal-base waveform design that enables controlled degradation of the wireless transmitted signal to preserve semantically significant content while minimizing resource consumption. 
We demonstrate that OSSDM not only reinforces semantic robustness against channel impairments but also improves semantic spectral efficiency by encoding meaningful information directly at the waveform level. Extensive numerical evaluations show that OSSDM outperforms conventional OFDM waveforms in spectral efficiency % energy efficiency 
and semantic fidelity. The proposed semantic waveform co-design opens new research frontiers for AI-native, intelligent communication systems by enabling meaning-aware physical signal construction through the direct encoding of semantics at the waveform level.
\end{abstract}
\smallskip

\vspace{.15 cm}
\section{Introduction }
\label{sec:intro}
The evolution of wireless communication systems toward 5G has primary focused on accurately transmitting each symbol, overlooking the inherent meaning of the source data and its ultimate pragmatic relevance for the goal at the destination \cite{CalvaneseGOWSC2021}. This \textit{transmit-without-understanding} design approach has stimulated the successful engineering of complex physical layer communication technologies, adaptive and reconfigurable antennas design and, the associated multi-parameter based resource allocation optimization mechanisms. 
This has enabled new levels of physical layer flexibility and reconfigurability, with an exploding number of configurable parameters and physical signals communication modalities, including multiple waveform configurations \cite{yazar2020waveform}. This complexity is further compounded by the potential need to support multiple waveform instances within the same frame, as for integrated communication and sensing (ISAC), to meet the simultaneous demands of different tasks, modalities or functions \cite{strinati2025toward}. Those trends have stimulated the support of Machine Learning (ML) and Artificial Intelligence (AI) to provide solutions in parameter assignments and optimization.

6G is offering a new playground for innovation toward \textit{network intelligentization} with the native integration of AI and the advent of semantic communication (SemCom) \cite{CalvaneseGOWSC2021}, enabling pervasive and collaborative intelligence across an extremely heterogeneous and distributed data sources and network infrastructure \cite{strinati2025toward}. This includes edge devices, sensing platforms, access points, and intelligent data planes. Recently introduced in the design of 6G \cite{CalvaneseGOWSC2021, chaccour2024less, uysal2022semantic}, %gunduz2022beyond}, 
SemCom leverages AI/ML to extract the minimum necessary information to transmit based on the meaning and its contextual pragmatic utility for a specific goal at the destination. This shift from physical signal optimized raw data transmission to goal-oriented, meaning-aware communication has the potential of drastically enhance efficiency and communication robustness, ensuring correct interpretation and pragmatic effectiveness of minimum informative transmission \cite{strinati2024goal, sana2022learning}. This also translates into a totally new opportunity of joint design of hardware, artificial intelligence and semantic-empowered effective goal-oriented communications \cite{strinati2021HWSemCom}.

The SemCom-aware physical layer design  stimulates one of the most dynamic and promising intersections in current 6G research, enabling new adaptation transmission strategies. The joint design of semantic and goal-oriented communication with physical layer functions, like multiple-input multiple-output (MIMO) \cite{weng2024semanticMIMO}, beamforming \cite{ma2024enhancedSemComRIS}, reconfigurable intelligent surfaces (RIS) \cite{hello2024SemComRISoptimiz} \cite{jiang2024SemComRIS}, ISAC \cite{strinati2025toward}, and interference management\cite{meng2024prompt} \cite{li2025semantic}, is an emergent open research area. While their full integration is still in its infancy, researchers has started to explore several promising directions.
Despite waveforms being fundamental to communication, their integration with SemCom remains underexplored. 

Current SemCom architectures often neglect waveform design, missing opportunities for joint optimization and semantic-aware encoding at the physical layer. This separation is increasingly inadequate. Instead, waveforms should be reimagined not as passive bit carriers but as active enablers of meaning-aware, goal-driven transmission. 
%A semantic waveform prioritizes interpretable information over bit-accuracy, optimizing for semantic spectral efficiency (SSE) \cite{yan2022resource}—measured by correctly interpreted meaning per Hz—shifting the focus from raw throughput to robustness against semantically irrelevant distortions.
A semantic waveform prioritizes interpretable information over bit-accuracy, shifting the focus from raw throughput to robustness against semantically irrelevant distortions.

Recent state of the art works have evaluated the combination of semantic extraction/interpretation modules with OFDM-based waveform \cite{yang2021deep}. The semantic gain has been validated but the waveform has been independently designed from the SemCom setting.
%Our observation is that, the traditional separation between waveform design and higher-layer semantic processing optimization becomes increasingly inadequate. Although waveforms form the physical foundation of any communication system, current SemCom architectures tend to operate independently of the waveform design, neglecting their joint performance optimization and overlooking the potential for semantic-aware physical layer encoding.
%%  Motivation for the paper
%Moreover, the increasing number of configurable waveform parameters, driven by new use cases (e.g., ISAC, XR, ultra-reliable low-latency communication), constraints on hardware capabilities, sustainability concerns and heterogeneous service requirements, makes the waveform design problem in 6G exponentially more complex. This complexity is further compounded by the potential need to support multiple waveform instances within the same frame, to meet the simultaneous demands of different tasks or modalities. 
%% 
To address those gaps, waveforms for SemComs must be rethought not as neutral carriers of bit-streams, but as active enablers of meaning-aware and goal-oriented informative information delivery.
Semantic waveform design involves selecting and configuring waveform parameters to maximize the effective semantic spectral efficiency (E-SSE)—defined not merely in bits per second per Hz, but in correctly interpreted semantic symbols per Hz per second at the receiver. %\cite{yan2022resource}. 
This implies a shift in design philosophy: from bit-accuracy to controlled degradation and robustness, where only semantically significant distortions matter.
%Moreover, the increasing number of configurable waveform parameters, driven by new use cases (e.g., ISAC, XR, ultra-reliable low-latency communication), constraints on hardware capabilities, sustainability concerns and heterogeneous service requirements, makes the waveform design problem in 6G exponentially more complex. This complexity is further compounded by the potential need to support multiple waveform instances within the same frame, to meet the simultaneous demands of different tasks or modalities. 

\noindent\textbf{Contribution:} We propose a novel semantic-aware waveform design—Orthogonal Semantic Sequency Division Multiplexing (OSSDM)—that leverages the inherent robustness of SemCom to reduce physical layer resource usage, %(power, spectrum, time), 
and benchmark its performance against classical OFDM-based waveforms. Our proposed parametrizable approach \cite{calvanese2024OSSDMPatent} trades controlled waveform degradation for sufficient semantic understanding at the receiver, optimizing resource use. OSSDM is built on the orthogonal Walsh basis, enabling energy-efficiency at the ADC/DAC levels compared to the used one in classical OFDM. The designed waveform carries semantic symbols containing semantically selected informative data through Walsh functions, directly shaping its representation in both the Walsh and time domains. To ensure compliance with system requirements, the selection of Walsh coefficients is restricted to respect hardware and RF chain constraints.
%\textcolor{purple}{The designed waveform carries semantic embeddings containing semantically selected informative data through Walsh functions, directly shaping its representation in both the Walsh and time domains. To ensure compliance with system requirements, the selection of Walsh coefficients is restricted to respect hardware and RF chain constraints} 
In fact, unlike traditional systems semantic encoding begins in our proposed framework at the waveform level, boosting E-SSE and showing substantial gains in semantic reconstruction, adding value beyond traditional source and channel coding. 
We introduce the \textit{Effective Semantic Spectral Efficiency (E-SSE)} metric that quantifies the rate at which semantic symbols are successfully interpreted at the receiver per unit of bandwidth and time over a noisy channel. Unlike traditional Spectral Efficiency (SE), which only operates at the physical level (Level A, \cite{Weaver1953}), E-SSE spans both the physical and semantic levels of communication (Levels A and B \cite{Weaver1953}). E-SSE specifically accounts for the correct semantic interpretation at the receiver, the influence of the channel quality, and the bandwidth and time resources consumed during transmission.
%E-SSE specifically accounts for the correct semantic interpretation at the receiver, the influence of the semantic channel quality (e.g., semantic alignment between sender and receiver), and the bandwidth and time resources consumed during transmission. 
Importantly, we propose computing this metric at the receiver side, which differentiates it from the Semantic Spectral Efficiency (SSE) metric proposed in \cite{Yan2022}, where SSE is evaluated at the transmitter based on the number of semantic messages sent per second over a given bandwidth. More details about the E-SSE concept are in end of section~\ref{sec:Numerical Results}.

\begin{figure*}[t]
\centering
\includegraphics[width=0.8\textwidth]{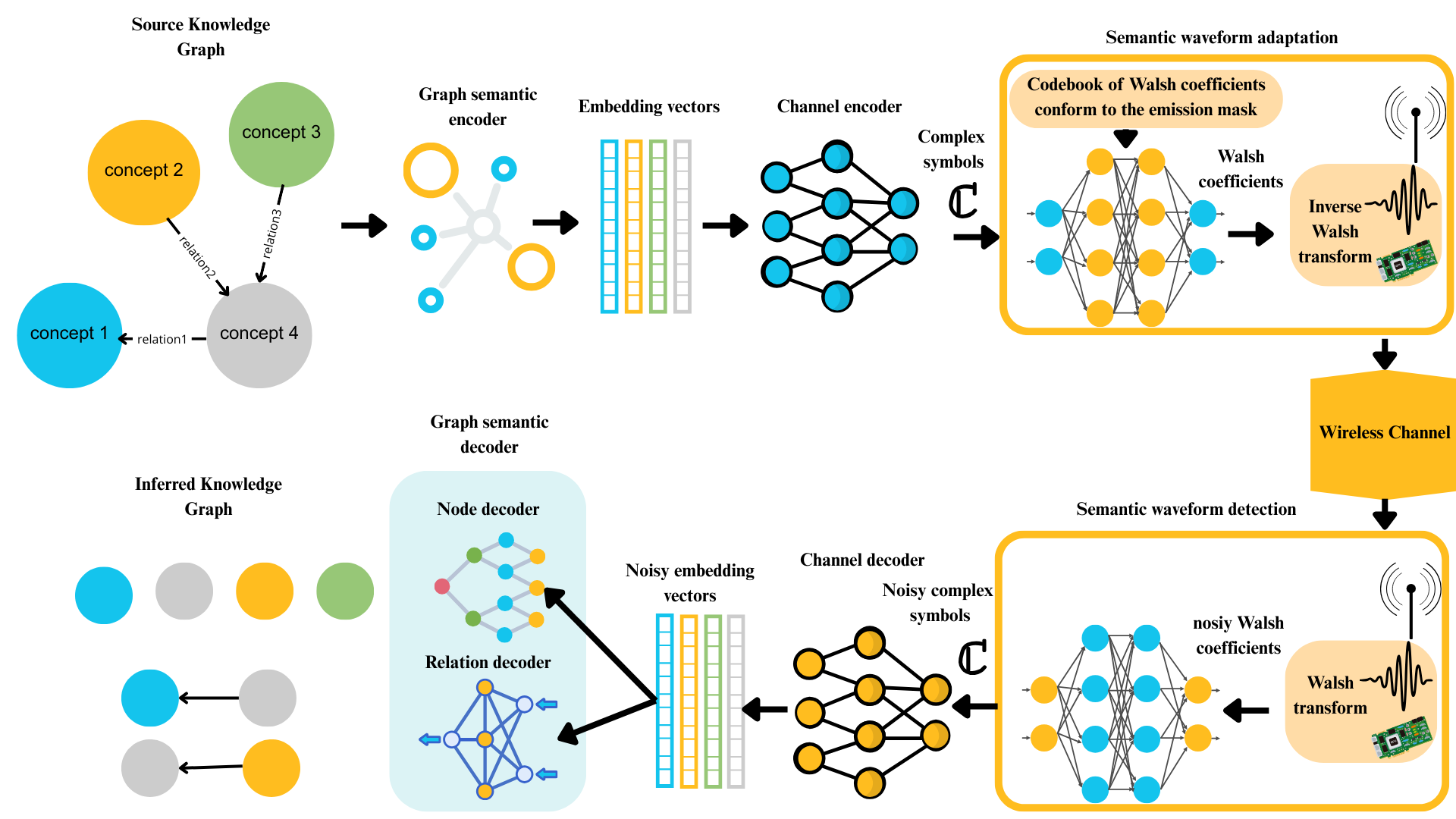}
\caption{System model and key components of the semantic-aware waveform communication chain.}
\label{fig:diag}
\end{figure*}

\section{System Model}
We consider the system model described in Fig.~\ref{fig:diag}, build upon the approach introduced in~\cite{Hello2024}. It leverages SemCom through knowledge graphs to enhance the efficiency of knowledge representation compression. We extend the SemCom model by incorporating a parametrizable waveform generation based on OSSDM. %% which leverages orthogonal basis representations for direct RF signal conversion %%. OSSDM exploits the properties of the Walsh matrix, a binary-valued orthogonal matrix with entries in \{-1, +1\}. In our approach, we utilize the Walsh Transform (WT) as a core operation to convert signals between the time domain and the \textit{Walsh domain}.
Let $\mathbf{x} \in \mathbb{R}^N$ be a signal of length $ N = 2^n$. The Walsh Transform (WT) of $\mathbf{x}$ is defined as:
\begin{equation}\label{WT}
\mathbf{a} = \mathbf{W}_N \mathbf{x}
\end{equation}
where $\mathbf{a}$ is the vector of walsh coefficients representing the signal $\mathbf{x}$ in the \textit{Walsh domain}, and $\mathbf{W}_N \in \mathbb{R}^{N \times N}$ is the Walsh matrix of order $n$. This matrix can be derived from the Hadamard matrix by reordering its rows according to increasing sequency—i.e., the number of zero-crossings in each Walsh function. We emphasize on the fact that the Walsh order n is among the design adaptable parameters.
%In the following subsections, we provide a detailed description of each component of our system model.
The detailed description of each key component in the proposed system model is provided below.

\noindent\textbf{Semantic Encoder:} The semantic encoder maps an input knowledge graph (KG) into a latent space representation. It consists of a cascaded architecture composed of a Large Language Model (LLM) encoder followed by a Graph Neural Network (GNN). The LLM associates feature vectors with the textual attributes of each concept node and relational edge in the KG. These features are then processed by the GNN, which to produce compact node-level embeddings that capture both concept and relational context. Then, a Feed Forward Neural Network (FFN) maps each node embedding from the latent space to a sequence of semantic complex symbols. In what follows, we denote the trainable parameters of the semantic encoder by $\theta$.
%The semantic encoding can be described as:
%\begin{equation}
%\mathbf{S} = f_{\theta}(\mathbf{g});
%\end{equation}
%where $f_{\theta}$ denotes the semantic encoder parameterized by $\theta$. Here, $\mathbf{g}$ represents the input knowledge graph, and $\mathbf{S}$ contains the complex semantic symbols associated with the graph nodes.

\noindent\textbf{Semantic Symbol to Walsh Coefficients Mapper:}
The Semantic Symbol to Walsh Coefficient Mapper predicts the corresponding Walsh coefficients to a given semantic symbol. This module takes as input a \textit{codebook} of Walsh coefficients, generated as we propose in Section~\ref{subsec:codebook}, to be compliant to the standard imposed emission mask (here defined by ETSI for the n53 band, specified for 5G base stations \cite{ETSI_TS_138_104_V16_7_0}). The goal of this module is to adapt the Walsh coefficients such that they are as close as possible to the nearest neighbor in the codebook, ensuring that the generated waveform complies with the emission mask. The mapping process is performed by an FFN parameterized by $\gamma$ which associates a vector of walsh coefficient of dimension $N=2^{n}$ with each semantic symbol.
Unlike OFDM, where complex symbols are directly transmitted, OSSDM requires a FFN to map between complex symbols and Walsh coefficients for waveform construction. This mapping is essential but does not independently improve performance over OFDM.
%The mapping process is defined by the following equation:
%\begin{equation}
  %  \mathbf{a} = q_{\gamma}(\mathbf{s});
%\end{equation}
%where $q_{\gamma}$ represents a neural network with trainable parameters $\gamma$, and $\mathbf{a} \in \mathbb{R}^{N}$ denotes the predicted Walsh coefficients. The dimension of $\mathbf{a}$ is $N=2^{n}$, where $n$ refers to the order of the Walsh basis, and $\mathbf{s}$ is a semantic complex symbol. 
%Nour: Added how mapping semantic symbols over a large base walsh generated being advantageous
This FFN adapts this mapping to generate physical signal representations that are inherently more resilient to noisy transmission environments. Notably, the dimensionality of this representation increases with higher Walsh orders $n$, contributing to improved robustness against noise.
%To ensure consistent signal power across different Walsh orders, the Walsh coefficient vector is normalized such that the resulting time-domain signal maintains an average power of $\frac{1}{64}$~watts, regardless of the Walsh order used.
\textcolor{green}{}
%\textcolor{purple}{As we know in conventional semantic communication systems, the semantic encoder already produces a latent represented designed to be robust to channel noise. As far as this is concerned, the added Walsh FFNs were introduced as a mandatory mapping that enables us to define our conceived Walsh-Based Semantic Waveform, by associating complex symbols to Walsh coefficients (and vice-versa), which is not needed for OFDM case since complex symbols are the ones to transmit. We emphasis on the fact that it does not bring an additional gain itself to the performance with respect to the OFDM structure, as it will be indeed demonstrated in the following sections.  }

%To achieve consistent normalization across different Walsh orders, the Walsh coefficient vector $\mathbf{a}$ is scaled as follows:
%\begin{equation}
 %  \mathbf{a}_{\text{norm}} = \frac{\mathbf{a}}{||\mathbf{a}||_2 \cdot \sqrt{2^{(\text{6} - n)}}}.
%\end{equation}
%This scaling ensures that when the Walsh order is $n = 6$, the resulting time-domain signal has unit energy, corresponding to a power of $\frac{1}{64}$~watts. 
%For Walsh orders ($n < 6$), the energy is reduced to $\frac{1}{2^{(6 - n)}}$, thereby maintaining the same average time signal power of $\frac{1}{64}$~watts across all orders.

\noindent\textbf{Waveform Generator:}
After the Walsh coefficient mapping, an Inverse Walsh Transform (IWT) layer reconstructs the discrete-time signal by summing the associations of Walsh coefficients and their respective Walsh functions:
\begin{equation}
\mathbf{x} = \mathbf{a}_{\text{norm}} \mathbf{W}_N^T;
\end{equation}
where $\mathbf{x}$ represents the reconstructed discrete temporal signal, $\mathbf{a}_{\text{norm}}$ is the normalized Walsh coefficients vector, and $\mathbf{W}_N$ is the Walsh matrix. 
Channel effects are simulated using either a Gaussian channel or a Rayleigh fading channel with additive white Gaussian noise.
%The resulting generated waveform integrates both physical and semantic aspects by embedding semantic information into the Walsh coefficients linked to the channel encoder’s complex symbols. 
%\textcolor{purple}{Consequently, the obtained Waveform is an immediate incorporation of both physical and semantic aspect, by carrying the semantic embedding of each symbol (obtained through the Walsh Coefficients that were semantically linked to the Complex Symbols of the Channel encoder). By receiving the Semantic Waveform as time-domain signal and using an Inverse Walsh transform, we perform a Mapping back to the semantic symbols and evaluate the reconstructed semantic messages (Text and Image related knowledge graphs). The performance results will be discussed in the following sections}

\noindent\textbf{Waveform to Walsh Coefficients:} At the receiver, the WT layer reconstructs the Walsh coefficients through \eqref{WT}.

%\begin{equation}
%\mathbf{\tilde{a}} = \mathbf{W}_N \mathbf{\tilde{x}};
%\end{equation}
%where $\mathbf{\tilde{x}} \in \mathbb{R}^{N}$ is the received discrete-time signal, $\mathbf{W}_N$ is the Walsh matrix, and $\mathbf{\tilde{a}}\in \mathbb{R}^{N}$ represents the decoded noisy Walsh coefficients.

\noindent\textbf{Walsh Coefficients to Semantic Symbol Mapper:}
Reconstructs the original complex symbol from the noisy Walsh coefficients of the received time signal. The reconstruction process is performed by an FFN, which takes the noisy Walsh coefficients as input and outputs the estimated semantic symbol. The FFN has trainable parameters $\eta$. 
%The reconstruction can be expressed as:
%\begin{equation}
%\mathbf{\tilde{s}} = v_{\eta}(\mathbf{\tilde{a}});
%\end{equation}
%where $v_{\eta}$ represents the feed forward neural network with the parameters $\eta$.

\noindent\textbf{Semantic Decoder:}
An FFN reconstructs latent vectors from the semantic symbols, which are then decoded
%Given $F$ complex symbols, an FFN architecture reconstructs the corresponding latent space vector. A total of $E$ such latent vectors are subsequently passed to the semantic decoder.The semantic decoder is designed 
with a branched architecture comprising a \textit{Node Decoder}, and a  \textit{Relation Decoder} that work in parallel to reconstruct the transmitted knowledge graph at the receiver. The semantic decoder's trainable parameters are denoted by $\phi$.
%\begin{equation}
%\mathbf{\tilde{g}} = g_{\phi}(\mathbf{\tilde{S}});
%\end{equation}
%where $g_{\phi}$ denotes the semantic decoder, parameterized by $\phi$. Here, $\mathbf{\tilde{S}}$ is the matrix of received complex symbols, and $\mathbf{\tilde{g}}$ is the inferred knowledge graph at the receiver.

\begin{figure}
    \centering
    \includegraphics[width=0.8\linewidth, height=0.6
    \linewidth]{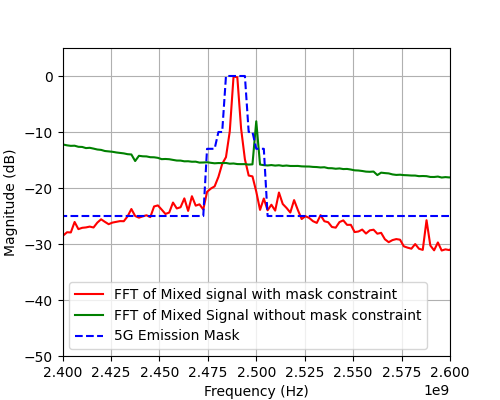}
   \caption{\centering{Semantic Waveform Spectrum w.r.t 5G Emission Mask \cite{ETSI_TS_138_104_V16_7_0}}.}
    \label{fig:spectrum}
\end{figure}

%Nour: Loss function from SPawc paper about Knowledge graph SemCom and we added to it a component for mask compliance
The end-to-end system is trained using a composite loss function that jointly optimizes concept and relation classification accuracy, while preserving semantic information, following a strategy similar to that in~\cite{Hello2024}. In addition, we incorporate a penalty term that enforces conformity to the spectral emission mask, ensuring that the generated waveforms complies with the mask constraints as shown in Fig.~\ref{fig:spectrum}.
%The total loss consists of three main components:
%\begin{enumerate}
 %   \item The sum of cross-entropy losses for both concept and relation classification tasks.
  %  \item The negative mutual information between the complex-valued symbols at the input of the Walsh mapping encoder and the output of the Walsh mapping decoder, encouraging information preservation across the communication channel.
   % \item The mean squared error (MSE) between the normalized predicted Walsh coefficients at the transmitter and their closest match in the predefined codebook.
%\end{enumerate}
The overall loss function is defined as follows:
\begin{comment}
\begin{equation}
\begin{aligned}
    \mathcal{L}_{\text{total}} =\ 
    &\mathcal{L}^{\text{CE}_\text{c}}_{\theta, \gamma, \eta, \phi} + 
    \mathcal{L}^{\text{CE}_\text{r}}_{\theta, \gamma, \eta, \phi} 
    - \alpha \, I_{\theta, \gamma, \eta}(\mathbf{S}; \tilde{\mathbf{S}}) 
    + \lambda \left\| \mathbf{a}_{\text{norm}} - \hat{\mathbf{c}} \right\|^2; \\
    &\text{where} \quad 
    \hat{\mathbf{c}} = \arg\min_{\mathbf{c} \in \mathcal{C}} 
    \left\| \mathbf{a}_{\text{norm}} - \mathbf{c} \right\|^2.
\end{aligned}
\end{equation}
\end{comment}
\begin{equation}
\begin{split}
\mathcal{L}_{\text{total}} =\ 
    & \mathcal{L}^{\text{CE}_\text{c}}_{\theta, \gamma, \eta, \phi} +
      \mathcal{L}^{\text{CE}_\text{r}}_{\theta, \gamma, \eta, \phi} \\
    & - \alpha \, I_{\theta, \gamma, \eta}(\mathbf{S}; \tilde{\mathbf{S}})
      + \lambda \left\| \mathbf{a}_{\text{norm}} - \hat{\mathbf{c}} \right\|^2 ;
\end{split}
\end{equation}
\begin{equation*}
    \text{where} \quad 
    \hat{\mathbf{c}} = \arg\min_{\mathbf{c} \in \mathcal{C}} 
    \left\| \mathbf{a}_{\text{norm}} - \mathbf{c} \right\|^2.
\end{equation*}

Here, $\mathcal{L}^{\text{CE}_\text{c}}_{\theta, \gamma, \eta, \phi}$ and $\mathcal{L}^{\text{CE}_\text{r}}_{\theta, \gamma, \eta, \phi}$ denote the cross-entropy losses for concept and relation predictions, respectively. The term $I_{\theta, \gamma, \eta}(\mathbf{S}; \tilde{\mathbf{S}})$ represents the mutual information between the semantic symbols $\mathbf{S}$ at the output of the channel encoder  and the noisy semantic symbols $\tilde{\mathbf{S}}$ at the input of the channel decoder . The coefficient $\alpha$ controls the contribution of the mutual information term, while $\lambda$ weights the codebook-based mean squared error penalty. The nearest codeword $\hat{\mathbf{c}}$ is selected from the codebook $\mathcal{C}$ based on the minimum Euclidean distance to the intermediate vector $\mathbf{a}_{\text{norm}}$.

\section{Proposed Semantic Waveform Design}
The Walsh codebook enables semantic symbol-to-waveform mapping under spectral and optimization constraints in the OSSDM transmitter design. 
In this section, we describe the rationale behind the codebook structure, its generation strategy, and its integration within the OSSDM training framework for the semantic mapping. We mention that 
each semantic symbol in OSSDM is mapped to a vector of Walsh coefficients that generate a time frame of an RF waveform. The time frame length is $\Delta T_{\text{SemSymbol}} = \frac{N}{F_{\text{refresh}}}$
\vspace{0.1cm}
such that $ {F_{\text{refresh}}} $ is the refresh rate of the Waveform Generator \cite{ferrer2023walsh}.
% Francois: j'ai mis $N$ en générique mais il faudrait peut être dire à un moment que $N=64$ car n=6 ? par la suite, je considère 64 comme acquis.

% Amine: J'ai ajouté les specifications comme dernière partie de cette section ou on mentionne les valeurs numériques pour notre cas: N = Ordre, B = Size du Block de Walsh, M = Nombre de lignes du codebook et L = Résolution du signal final. 
\textbf{Codebook Construction and Sampling Strategy}:\label{subsec:codebook} To overcome the challenge of generating semantic-aligned waveforms while ensuring spectral conformity, we propose a structured Walsh-based codebook generation strategy. 
Let each codeword $\mathbf{c}^{(i)} \in \mathbb{R}^{L}$, $i \in \{1,\dots,M\}$, represent a real-valued waveform composed of $B$ Walsh blocks of length $N$, such that $L = B \cdot N$. Each Walsh block is generated by projecting a localized fragment of a continuous sinewave signal, denoted as $\mathbf{x}^{(i)}_b$, onto an orthogonal Walsh basis $\mathbf{W}_N \in \mathbb{R}^{N \times N}$, yielding a Walsh coefficient vector $\mathbf{\hat{x}}^{(i)}_{b} = \mathbf{W}_N \mathbf{x}^{(i)}_b$ for block $b \in \{1,\dots,B\}$.
%($W_N$ is the Walsh matrix of order $N$). 
This formulation leverages the blockwise orthogonality of the WT to preserve time-frequency localization and facilitate independent learning.
The codebook $\mathcal{C} = \{\mathbf{c}^{(i)}\}_{i=1}^M$ is populated using sampled sinewaves spanning a tightly controlled frequency range of emission mask. Each codeword encodes a distinct frequency component, with blockwise Walsh coefficients modulated via a perturbation mechanism to enforce power variability and generalization across training. Specifically, we inject a Gaussian perturbation on the target power for each block with mean $\mu = \frac{1}{B}$ and small variance $\sigma^2$, ensuring energy diversity across codebook entries.
%
% Francois: I do not understand, what is W ? clarify.
% Amine: It meant the unit power in Watts, but I think it is irrelevant so it is removed ...
During reconstruction, one Walsh-encoded block is selected per segment from potentially distinct rows of the codebook (but still respecting the ascendant time domain order). This random blockwise assembly yields a final waveform that approximates a coherent target sinewave-like signal while preserving diversity and block-level independence. The block structure aligns naturally with the Walsh basis and facilitates training of OSSDM, which requires localized updates.
\begin{comment}

\textcolor{red}{Is is feasible to add such a figure to both explain the Codebook Based training and also how the Semantic Waveform is generated ?}

\begin{figure}[H]
    \centering
    \includegraphics[width=0.9\linewidth, height=0.5
    \linewidth]{figures/codebook_trainnig.png}
    \caption{Codebook based training and generated Semantic Waveform}
    \label{fig:f1score}
\end{figure}

\end{comment}
%
% Francois: pas clair ce qu'est un Walsh block: to be clarified. 

% Amine: J'ai ajouté une clarification entre parenthèese pour cela. 
\textit{Numerical specification.} For the evaluation in section~\ref{sec:Numerical Results}, we set $N = 64$ and $B = 64$, yielding $L = 4096$ samples per waveform. The mean deviation is set to $\sigma^2 = \frac{0.25}{B}$ and the codebook is instantiated with $M = 10^4$ entries. %These settings correspond to the numerical resolution and are not required by the general methodology described above.
%
% Francois: this part is again a choice that should be explicated at the numerical evaluation section. On the other hand how the codebook is built this belongs to this section.

\section{Numerical Results}
\begin{figure*}[htbp]
    \centering
    \subfigure[AWGN Channel]{
        \includegraphics[width=0.4\textwidth]{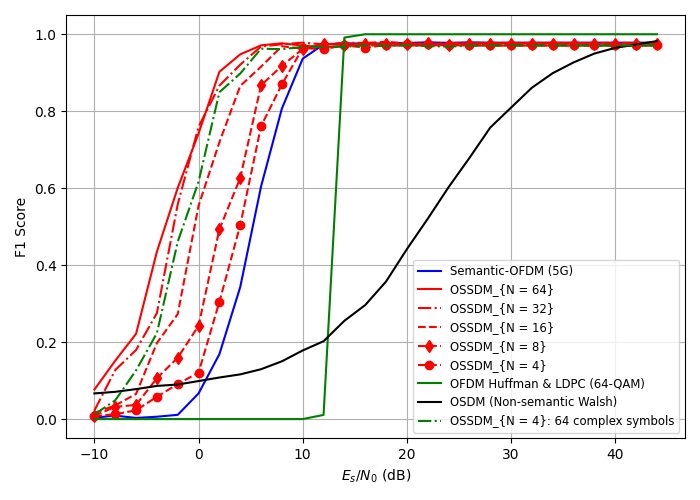}
        \label{fig:awgn}
    }
    \hfil
    \subfigure[Rayleigh fading Channel]{
        \includegraphics[width=0.4\textwidth]{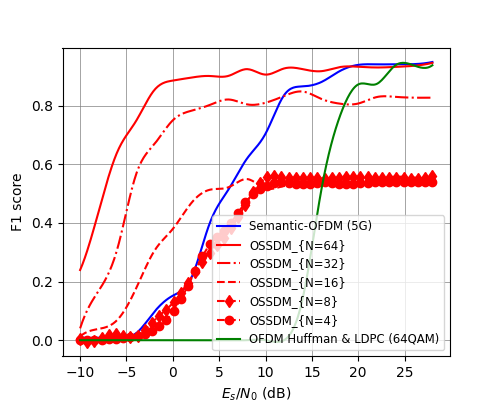}
        \label{fig:rayleigh}
    }
    
    \caption{\centering{Performance comparison of OSSDM with OFDM-based baselines, including (i) non-semantic OFDM with LDPC channel coding and (ii) semantic symbol transmission over conventional OFDM.}}
    \label{fig:f1score}
\end{figure*}
\label{sec:Numerical Results}
%For each selected Walsh order, a distinct instance of the proposed system model was trained and evaluated by configuring the corresponding Walsh matrix for both the forward and inverse Walsh transforms. 
%To ensure energy preservation, the Walsh matrix was normalized by a factor of $\sqrt{2^n}$, where $n$ denotes the Walsh order. This scaling guarantees that the energy of the signal obtained after the inverse Walsh transform remains consistent with that of the original Walsh domain coefficients. Additionally, the corresponding codebook specific to the selected Walsh order was configured to synthesize OSSDM waveforms compliant with the ETSI spectral emission mask for 5G band n53.
%On the semantic encoder side, we employ the pre-trained language model ‘all-MiniLM-L12-v2’ \cite{reimers-2019-sentence-bert} to extract 384-dimensional feature representations for the concepts and relationships within a given knowledge graph. We implement the graph neural network using the Graph Isomorphism Network (GIN) architecture, as described in \cite{gine}. 
The simulation settings used in this work closely follow those presented in \cite{Hello2024} except for the following simulation details:
%The GNN leverages the graph's relational topology to project node-level features into a lower-dimensional latent space, producing embeddings of size 16. These 16-dimensional vectors are subsequently converted into four complex-valued symbols through a channel encoder. 
the end-to-end SemCom models are trained on the WebNLG dataset of knowledge graphs, %specifically on graphs composed of seven triplets, 
under a signal-to-noise ratio (SNR) of 14 dB and with a mini-batch size of two graphs. The 16-dimensional latent space embeddings generated by the GNN are modulated into four complex-valued semantic symbols. Within the training loss function, the weighting coefficients are configured with $\alpha=0.1$, and $\lambda=1$.

\textbf{Impact of Waveform Design on Semantic Preservation in Noisy Environments:}
to evaluate the preservation of semantic information under low-SNR conditions, we compared the performance of the proposed OSSDM waveform against both semantic and traditional OFDM-based transmission schemes.
In both OFDM-based transmission schemes, complex symbols were transmitted using OFDM parameters specified in the 5G NR standard for Frequency Range 1 (FR1) \cite{ETSI_TS_138_104_V16_7_0}, with a subcarrier spacing (SCS) of 30~kHz and a total bandwidth of 10~MHz, corresponding to 24 resource blocks and a cyclic prefix duration of 2.3~$\mu$s.
For the \textit{Semantic OFDM} baseline (S-OFDM), this comparison was conducted by replacing the semantic waveform adaptation and detection modules in our system model with conventional OFDM-based modules.  %In this scenario, to ensure a fair comparison with OSSDM, the transmit power allocated per semantic symbol was fixed at $\frac{1}{64}$~watts. 
To ensure a fair comparison and consistent signal power, the transmit power per semantic symbol was fixed at $\frac{1}{64}$~watts in both OSSDM (across different orders) and S-OFDM.
%To isolate the performance gains of OSSDM from the influence of additional neural components (i.e., the semantic symbol-to-Walsh coefficients mapper and demapper), the same FFNs used in OSSDM were also applied to the S-OFDM baseline. 
 Additionally, a \textit{Non-Semantic} OFDM baseline was evaluated, where the knowledge graph (KG) was first compressed using Huffman source encoding, followed by LDPC channel coding with a rate of 0.5, 64-QAM modulation, and OFDM transmission. 
 For each Walsh order $n$, a distinct OSSDM system model was trained using the corresponding normalized Walsh matrix by a factor of $\sqrt{2^n}$, to ensure energy preservation between the \textit{Walsh domain} and the \textit{time domain}. Compliance with the ETSI emission mask was also ensured.
 To further assess the contribution of semantic coding, we introduce a non-semantic Walsh-based transmission baseline, referred to as OSDM. In this setup, each KG triplet (concept-relationship-concept) is randomly mapped to a vector of 64 Walsh coefficients using the predefined codebook, without any semantic-aware coding or learning. At the receiver, reconstruction is performed by applying the WT, 
followed by nearest-neighbor decoding based on the minimum Euclidean distance to the original codebook entries.
%\textcolor{red}{On the other hand, we introduce a non-semantic OSSDM (OSDM) as a Walsh-Based Transmission relying only on the a triplet-to-walsh-vectors Codebook Mapping. The reconstructed vectors after channel transmission at the Reception end (using only the Walsh Transform without Semantic Demapping/Decoding) are decoded through the minimal distance to the reference codebook, highlighting the non semantic transmission/reception aspect}. 
In our experiments, the F1 score provides a comprehensive measure of how well the information and structure of the source KG are preserved in the inferred KG, serving as a key indicator of the semantic fidelity of the proposed system.
Specifically, the F1 score quantifies the balance between precision and recall when identifying correct triplets in the graphs, %The F1 score is calculated as:\[
%F1 = 2 \times \frac{Precision \times Recall}{Precision + Recall};
%\]
where precision is the proportion of correctly identified triplets in the inferred KG out of all identified triplets, and recall is the proportion of correctly identified triplets in the inferred KG out of all triplets present in the source KG.

\begin{figure}
    \centering
    \includegraphics[width=0.8\linewidth, height=0.6
    \linewidth]{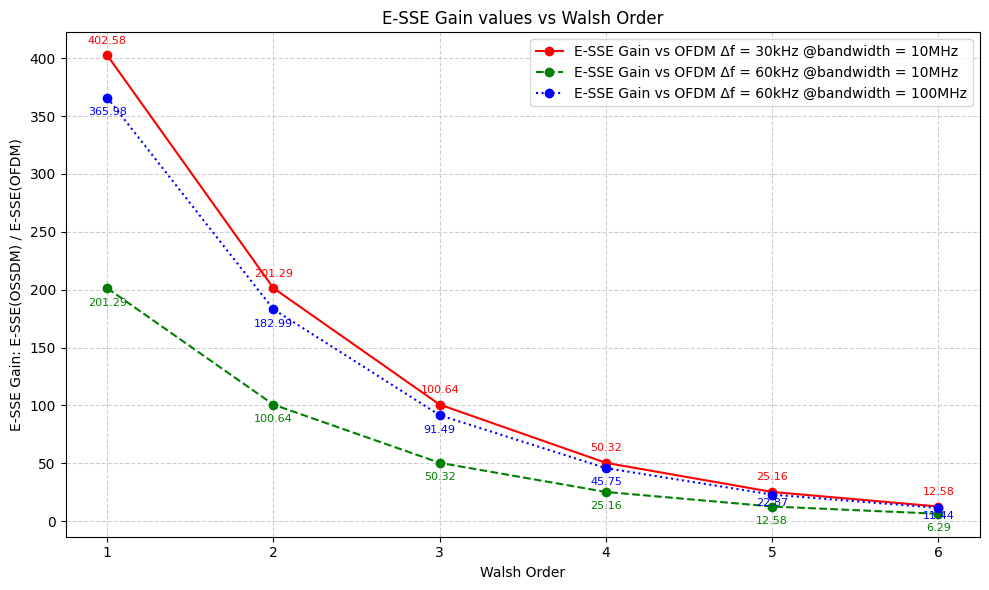}
    \caption{\centering{E-SSE Gain of OSSDM w.r.t different OFDM numerologies, at maximal F1 score.}}
    \label{fig:esse_gain}
\end{figure}

%The F1 score ranges from 0 to 1, where 1 represents perfect precision and recall (i.e., the inferred KG exactly matches the source KG), and 0 indicates no overlap between the two graphs.

Fig.~\ref{fig:awgn} demonstrates that over additive white guaussian noise (AWGN) channel conditions, OSSDM with  $N=64$ achieves the maximum F1 score with an SNR that is 6dB lower than that required by the S-OFDM baseline. A similar performance gain is observed for OSSDM with $N=32$. For OSSDM with $N=16$ and $N=8$, the SNR gains with respect to S-OFDM are 4 dBs and 2 dBs, respectively, while OSSDM with $N=4$ reaches its maximum F1 score at the same SNR as the S-OFDM scheme. 
These improvements are attributed to the learnable nature of the OSSDM waveform generation process. 
While both OSSDM and S-OFDM enable the transmission of semantic symbols, S-OFDM relies on  
OFDM-specific operations such as Fast Fourier Transform (FFT) and its inverse (IFFT), which are not inherently adapted to semantic coding. In contrast, OSSDM integrates a trainable mapping from semantic symbols to \textit{Walsh domain} representations and leverages the WT and IWT, which are simple matrix multiplications. This simplicity not only reduces computational complexity but also aligns more naturally with neural architectures, making it more compatible with semantic communication paradigms.

Furthermore, despite augmenting the S-OFDM baseline with the same FFN blocks used in OSSDM, 
no significant performance gain was observed. This highlights that the advantage of OSSDM stems not merely from the neural components but from the end-to-end learnability of its waveform structure, which is inherently more aligned with the neural architecture of the semantic coding than the OFDM modules.
%These improvements are attributed to the learnable nature of the OSSDM waveform generation process. While both OSSDM and S-OFDM enable the transmission of semantic symbols, OFDM relies on a fixed modulation and does not adapt the waveform to the underlying semantic content. In contrast, OSSDM integrates a trainable mapping from semantic symbols to \textit{Walsh domain} representations, allowing the model to optimize waveform characteristics to improve robustness under noisy channel conditions. 
Additionally, OSSDM with $N=64$ requires up to 10 dBs less SNR than traditional \textit{Non-Semantic} OFDM, demonstrating the benefits of semantic-aware waveform design. On the other hand, The performance of OSDM was found to be significantly worse than that of the \textit{Non-Semantic} OFDM scheme. 
This is due to that OSDM uses a fixed, random mapping from KG triplets to Walsh coefficient vectors without incorporating any channel coding. 
Since each triplet is mapped arbitrarily to a Walsh vector and no error correction is employed, the transmission is highly vulnerable to noise. Moreover, the receiver's reliance on simple nearest-neighbor decoding in the Walsh domain further limits robustness. 
%especially in the absence of adaptive mechanisms or learned representations. 
These results underscore the critical role of semantic coding and learning-based waveform adaptation in achieving robust and efficient transmission of KGs. To isolate the effect of representational capacity from transmission duration, we conducted a control experiment matching the transmission time per knowledge graph concept node by comparing OSSDM (64) with 4 complex symbols per concept node, to OSSDM (4) setting 64 complex symbols per concept node. 
Consequently, as shown in Fig.~\ref{fig:awgn}, the higher order system still outperforms the lower one, confirming that the performance gain stems from increased mapping diversity and expressiveness in the Walsh domain.
%\textcolor{purple}{To ensure that the performance gains of higher Walsh orders are not primarily due to longer transmission durations, we conducted a control experiment comparing OSSDM at $N=64$ (using the default 4-complex symbol representation) to a specific OSSDM configuration at $N=4$ (OSSDM(4): using 64 complex symbols), ensuring the same time transmission per semantic symbol. As shown in Fig.~\ref{fig:f1scoreawgn}, the higher-order system still outperforms the lower one, confirming that the gain arises from greater representational capacity and enhanced mapping diversity in the Walsh domain.}

Fig.~\ref{fig:rayleigh} demonstrates that under the 3GPP Tapped Delay Line B (TDL-B) channel model, which captures Rayleigh fading and multipath effects, S-OFDM significantly outperforms OSSDM variants with $N = 16$, $N = 8$, and $N = 4$ when Minimum Mean Square Error (MMSE) equalization is applied for multipath compensation. Specifically, S-OFDM achieves F1 scores greater than 0.8 at SNR levels exceeding 10~dB, due to its ability to exploit frequency diversity through subcarrier separation, which enhances its robustness against fading. In contrast, OSSDM maps each semantic symbol to $N$ Walsh coefficients, and the IWT generates a time-domain signal that lacks frequency diversity. As $N$ decreases (from 32 to 4), OSSDM performance degrades due to reduced signal dimensionality, making it more susceptible to fading effects. 
%These findings suggest that while OSSDM benefits from orthogonal spreading, it lacks mechanisms such as OFDM’s subcarrier diversity that are critical in fading environments.
However, OSSDM with $N = 64$, can outperform S-OFDM particularly at low SNR. This is because OSSDM distributes the energy of each semantic symbol across multiple Walsh coefficients and transforms them into a full-length time-domain signal via IWT. Each time sample carries partial information about the symbol, providing implicit redundancy. As a result, even when some samples are corrupted by noise, the remaining ones still contribute to accurate symbol recovery.

\textbf{Effective Semantic Spectral Efficiency (E-SSE) Analysis:}
To quantify the transmission efficiency of the proposed OSSDM system compared to classical OFDM, we analyze the E-SSE gains across different Walsh orders. 
\textbf{E-SSE} reflects the real effectiveness of communication. It acknowledges that, even with identical physical channel conditions and equivalent semantic extraction/compression strategies, the actual communication success—measured at the receiver—depends critically on the semantic channel quality. For instance, severe semantic misalignment between transmitter and receiver (e.g., incompatible ontologies or context models) can cause E-SSE to drop to zero, even if the physical channel offers high SNR and classical SE is maximized. In such cases, the semantic channel is effectively closed, and meaningful communication fails despite ideal physical conditions. E-SSE is defined as:
\[
\text{E-SSE} = \frac{|\mathcal{S}^*|}{B \times T} \quad [\text{symbols/s/Hz}]
\]
where \( |\mathcal{S}^*| \) is the number of semantic symbols successfully interpreted at the receiver, \( B \) denotes the channel bandwidth in Hertz (Hz), and \( T \) is the duration of the communication interval in seconds. The set \( \mathcal{S} \) represents the transmitted semantic symbols, and \( \mathcal{S}^* \subseteq \mathcal{S} \) captures those correctly decoded at the receiver. The semantic symbol rate is thus given by \( R_s = \frac{|\mathcal{S}^*|}{T} \), reflecting the true end-to-end throughput in semantically meaningful units.
\begin{comment}

Formally, let:

$\bullet$ $B$ be the channel bandwidth in Hertz (Hz) \\
$\bullet$ $T$ be the duration of the communication interval (in seconds) \\
$\bullet$ $\mathcal{S} = \{s_1, s_2, \dots, s_n\}$ be the set of semantic symbols transmitted \\
$\bullet$ $\mathcal{S}^* \subseteq \mathcal{S}$ be the subset of semantic symbols successfully interpreted at the receiver \\
$\bullet$ $|\mathcal{S}^*|$ denote the number of correctly interpreted semantic symbols \\
$\bullet$ $R_s = \frac{|\mathcal{S}^*|}{T}$ be the semantic symbol rate (symbols/second)

\vspace{0.2cm}

The \textbf{E-SSE} is then defined as:

\[
\text{E-SSE} = \frac{|\mathcal{S}^*|}{B \times T} \quad [\text{symbols/s/Hz}]
\]
%}
\end{comment}
The E-SSE Gain results are shown in Fig.~\ref{fig:esse_gain}, comparing OSSDM to multiple OFDM configurations across Walsh orders from $1$ to $6$. The gain is computed as the ratio between the E-SSE of OSSDM and that of OFDM, i.e., $\text{E-SSE}_\text{OSSDM} / \text{E-SSE}_\text{OFDM}$.
Three OFDM baselines are considered: a standard configuration with $\Delta f = 30$\,kHz at 10\,MHz bandwidth, a second with $\Delta f = 60$\,kHz at 10\,MHz, and a third with $\Delta f = 60$\,kHz at 100\,MHz. Across all cases, the E-SSE gain decreases with increasing OSSDM order, due to the reduced symbol density in higher-order transforms.
At order $1$, OSSDM provides a gain exceeding $400\times$ over OFDM with $\Delta f = 30$\,kHz, while at order $6$, the gain converges below $13\times$ across all OFDM variants. The most significant differences are observed at lower orders, where OSSDM encodes a larger number of semantic symbols in shorter waveform durations. These results highlight the capacity of OSSDM to adapt spectral efficiency via the parameterized Walsh order control, outperforming OFDM across a wide range of physical layer constraints.

\section{Conclusions}
Existing semantic communication architectures treat physical-world waveform design as a separate concern from logical-level semantic processing and the goal of the communication. This separation is increasingly inadequate for the demands of future 6G networks, which require tight integration between physical layer efficiency and goal-oriented semantic-level robustness. In this paper, we propose Orthogonal Semantic Sequency Division Multiplexing (OSSDM), a novel semantic-aware waveform that adaptively encodes meaning directly into the transmitted waveform using Walsh basis functions. OSSDM introduces a parametrizable waveform structure that allows controlled signal degradation to preserve semantically relevant content, while also minimizing %power
bandwidth and hardware resource usage and meeting system constraints (spectrum, energy).
Unlike conventional OFDM, which carries blindly complex symbols without semantic awareness, OSSDM jointly optimizes %communication robustness and 
semantic robustness, by integrating physical and semantic layers at the signal level. Additionally, OSSDM adapts to hardware constraints, making it suitable for energy-efficient RF architectures and compatible with implementation-specific digital-to-analog converters. 
Our simulations demonstrate that OSSDM consistently outperforms both standard LDPC-coded OFDM and semantic OFDM baselines in terms of semantic fidelity %energy efficiency 
and effective spectral utilization. These results confirm the value of OSSDM as a key enabler for AI-native, goal-oriented wireless systems. By introducing a co-designed waveform that embeds meaning into the physical signal itself, OSSDM opens a new research direction where AI, semantics, signal processing, and hardware converge in the design of next-generation intelligent communication systems.

\bibliographystyle{IEEEtran}
\bibliography{bibliography}
\end{document}